# Thousand-fold Increase in Plasmonic Light Emission via Combined Electronic and Optical Excitations

*Longji Cui[1,2,3,†], Yunxuan Zhu[1,†], Peter Nordlander[1,4,5], Massimiliano Di Ventra[6], Douglas Natelson[1,4,5,*]*

[1]Department of Physics and Astronomy and Smalley-Curl Institute, Rice University, Houston, TX 77005, United States.

[2]Paul M. Rady Department of Mechanical Engineering, University of Colorado, Boulder, CO 80309, United States.

[3]Materials Science and Engineering Program, University of Colorado, Boulder, CO 80309, United States.

[4]Department of Electrical and Computer Engineering, Rice University, Houston, TX 77005, United States.

[5]Department of Materials Science and Nanoengineering, Rice University, Houston, TX 77005, United States.

[6]Department of Physics, University of California San Diego, La Jolla, CA 92093, United States.

[†]These authors contributed equally to this work.

[*]Corresponding author: Douglas Natelson (natelson@rice.edu)

KEYWORDS: Plasmonics, light emission, tunnel junction, hot-carrier dynamics, synergistic effect

**ABSTRACT**

Surface plasmon enhanced processes and hot-carrier dynamics in plasmonic nanostructures are of great fundamental interest to reveal light-matter interactions at the nanoscale. Using plasmonic tunnel junctions as a platform supporting both electrically- and optically excited localized surface plasmons, we report a much greater (over 1000×) plasmonic light emission at upconverted photon energies under combined electro-optical excitation, compared with electrical or optical excitation separately. Two mechanisms compatible with the form of the observed spectra are interactions of plasmon-induced hot carriers and electronic anti-Stokes Raman scattering. Our measurement results are in excellent agreement with a theoretical model combining electro-optical generation of hot carriers through non-radiative plasmon excitation and hot-carrier relaxation. We also discuss the challenge of distinguishing relative contributions of hot carrier emission and the anti-Stokes electronic Raman process. This observed increase in above-threshold emission in plasmonic systems may open avenues in on-chip nanophotonic switching and hot carrier photocatalysis.

**Introduction**

Optically excited localized surface plasmons (LSPs) in metal nanostructures have been studied extensively and hold promise for technologies including surface-enhanced spectroscopies[1,2], photosensing[3,4], photocatalysis[5–7] and photovoltaics[8]. In recent years progress has also been made in understanding electrically generated LSPs and optical nanoantenna effects [9–11], opening the potential for plasmonic applications controlled by electronic means. Plasmonic tunnel junctions emerge as a unique experimental platform that supports LSPs generated by both electrical and optical excitations[10–15], strongly confined in an ultra-small nanogap. In electrically driven tunnel junctions, electrons tunneling from source to drain inelastically excite LSPs[16,17], which subsequently undergo rapid relaxation via radiative or non-radiative decay.

Recent studies[18–23] of plasmonic light emission in tunnel junctions have reported a strong upconversion effect, with generated photon energy ($\hbar\omega$) significantly above the energy threshold of the incident electrons ($eV$ with $V$ the applied bias). While multiple processes can produce such above-threshold photons, recent work[24,25] has revealed the dominant role of LSP-induced hot carriers. Similarly, upconversion photoluminescence ($\hbar\omega > \hbar\omega_{exc}$ with $\hbar\omega_{exc}$ the excitation photon energy) from plasmonic nanoparticles[26–29] has also been observed, and this phenomenon can be explained by mechanisms such as hot carrier luminescence[26], anti-Stokes electron Raman scattering[27,30,31], and other surface-enhanced anti-Stokes processes[32–34]. While previous studies focused on driving plasmonic nanostructures using either electrical or optical excitation alone and provided insight into the relevant hot-carrier dynamics, the effects of multiple excitation sources remain less explored. A recent work[35] has reported that under electrical and optical excitations, light emission from a Au junction is enhanced as much as six-fold (total photon counts divided by the simple sum of photons emitted under each excitation). However, such enhancement is driven by the optical interband transition of Au, which only generates below-threshold photons (i.e., photon energy is below the excitation energy). These findings raise the question: Is it possible to excite the system via concurrently applied electronic and optical drive into a regime such that upconversion processes dominate the plasmonic response?

In this work, we report a large increase in upconverted photon emission when simultaneous optical and electrical excitations are applied to plasmonic junctions. In Au devices, we found that electrically biased, optically pumped junctions emit over 1000× more upconverted photons than the simple sum of emission under either electrical or optical stimulus alone (Fig. 1A), demonstrating a broadband plasmonic switchable light source controlled by electrical voltage or input optical power. This increase is not just a case of dividing by a small denominator; in the electrically and optically pumped junctions, the above-threshold photons are a majority of the total emission. Analysis of the emission spectra with and without optical pumping reveals that the possible mechanisms involved in this increase are 1) joint production of hot carriers, manifested microscopically as an increase in the effective temperature of the steady-state hot carriers with the addition of optical excitation, and 2) anti-Stokes electronic Raman scattering, with the applied bias and local wavefunctions setting the phase space for Raman processes.

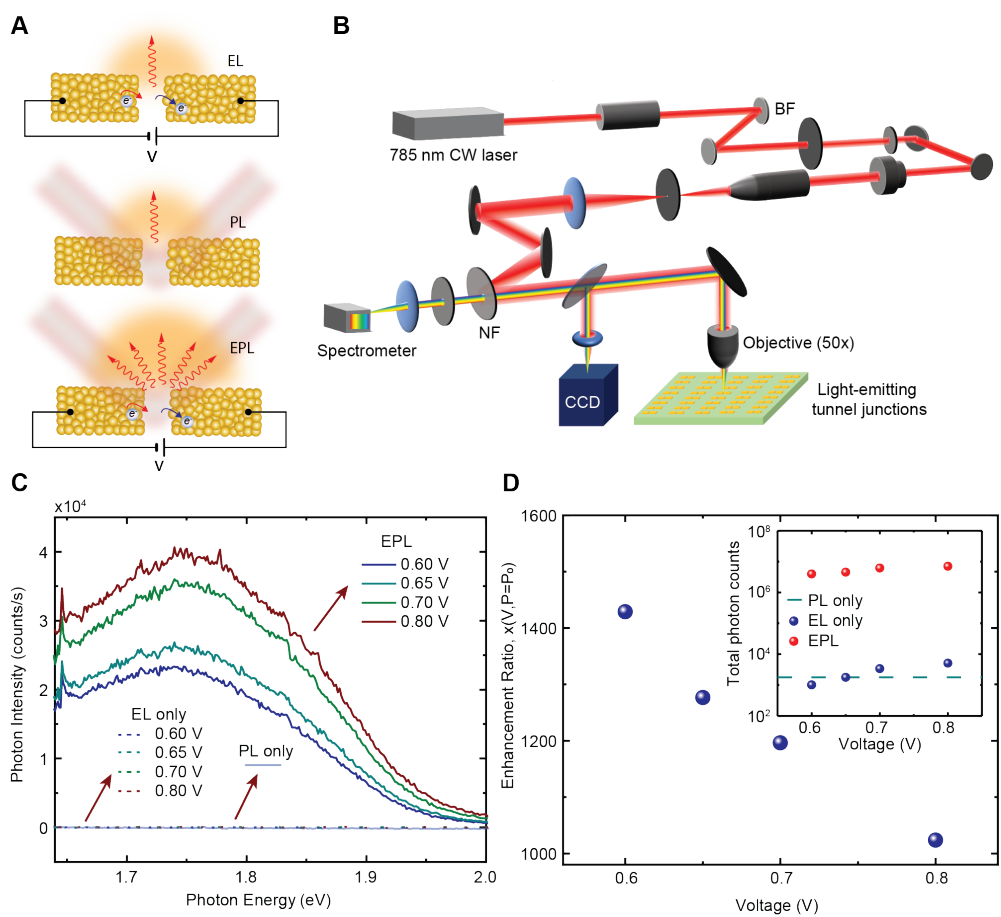

**Figure 1.** Experimental strategy and observation of giant increase in upconversion light emission from plasmonic tunnel junctions. (**A**) Schematics of experimental design to measure light emission under three types of external stimuli (EL, electroluminescence; PL, photoluminescence; EPL, electro-photo-luminescence). (**B**) Schematics of the combined electrical-optical setup. BF, Bragg filter; NF, Notch filter. The diameter of the focused laser beam through the 50× objective is 1.8 µm. (**C**) Spectral emission intensity of EL, PL, and EPL for photon energy larger than the energies of both the laser photons and tunneling electrons in a Au junction. The small peaks on EPL between 1000-1800 cm$^{-1}$ (~1.7-1.8 eV) are anti-Stokes SERS emission of residual contaminant organic molecules on the sample. The applied laser power in PL and EPL is 0.345 mW. (**D**) Measured enhancement ratio (total upconverted photons in EPL divided by the sum of EL and PL measured in **C** vs. the applied bias with fixed incident light power 0.345 mW. Inset shows the total photon counts for EL, PL, and EPL, respectively, at different biases. The photon count of PL is indicated by the dash line.

**Experimental results**

The combined optoelectronic experimental setup is shown in Fig. 1B. We fabricated tunnel junctions from arrays of Au nanowires and obtained tunnel junctions by employing the electromigration break junction (EBJ) technique. (see Supplementary Information Sec. 1 and 2 for the nanofabrication and electromigration procedure). Subsequent to the creation of the tunneling gaps, light emission measurements were conducted on each junction under three conditions (voltage bias with no incident light; continuous wave (CW) optical excitation at 785 nm with no voltage bias; and optical excitation in the presence of voltage bias), producing electroluminescence (EL), photoluminescence (PL), and electro-photo-luminescence (EPL), respectively (Fig. 1A).

Figure 1C shows the light emission spectra ($\hbar\omega > \hbar\omega_{exc} \approx 1.58\ eV$) measured from a Au junction. In the EL case, above-threshold emission can be generated via multi-electron interactions in the low current limit (~100 nA)[19], and hot carrier recombination in the high current limit (~100 µA)[24]. For PL at zero bias, in addition to the hot carrier[26] or intraband transition[36] induced photoluminescence, tunneling electrons can also undergo a Raman process (electronic Raman scattering, ERS) before emitting a photon with a different energy[30,31]. As shown in the discussion below, the anti-Stokes emission for the unbiased ERS is mainly due to the thermally excited tail of the electron-hole joint distribution and density of states[31] and thus cannot extend to large wavenumbers beyond $k_B T$. Hence both EL at low bias and PL can only emit few photons at upconversion photon energies, as indicated in Fig. 1C. Surprisingly, the junction under simultaneous optical and electrical excitation (EPL) emits far more light than when driven at the individual EL and PL excitation. In the following discussion, we define the enhancement ratio

$$x(V, P) = \int_{1.65\ eV}^{2\ eV} U^{EPL}(\omega, V, P)d\omega \Big/ \int_{1.65\ eV}^{2\ eV} [U^{EL}(\omega, V, P = 0) + U^{PL}(\omega, V = 0, P)]d\omega \quad (1)$$

where $P$ is the optical power and 2.0 eV is chosen as the upper bound of the integral since little light was seen emitted from the junction above this energy, corresponding to the interband transition of Au. As shown in Fig. 1D, EPL generates over 1000 times more photons than the total amount of photons generated by the sum of EL and PL. This is a dramatic demonstration that electrical and optical excitation work in cooperation in the upconversion emission process. To verify the reproducibility of this increase in emission, we measured in total 14 pure Au junctions.

Since multiple mechanisms could be contributing to the observed EPL, we will first focus on presenting the experimental results of EPL under different experimental conditions and then discuss the underlying microscopic processes. We performed measurements and analysis of the upconversion emission spectra, the enhancement ratio $x(V, P)$, and their dependence on the applied voltage, laser power, and plasmonic materials. Figure 2 shows the results for a Au junction under different biases without (Fig. 2A) and with (Fig. 2D) incident power at 0.46 mW. Our past work[24] on light emission from electrically driven junctions shows that, using a normalization method, one can separate the voltage-independent LSP spectrum, $\rho(\omega)$. Here we perform a similar analysis (see Supporting Information Sec. 6 for details) on the spectra in Fig. 2. We first analyze the EL results in the high current limit (Fig. 2A, ~20 μA) by normalizing the measured spectra at 0.75 V, 0.80 V, and 0.85 V to the spectrum at 0.90 V. Figure 2B shows the reduced spectra, plotted on log scale, in which the normalized log spectral intensity linearly decreases with the photon energy. In our previous work[24], we used a voltage dependent effective temperature $T_{eff}$ in a Boltzmann factor, $e^{-\hbar\omega/k_B T_{eff}}$, to parameterize this exponentially decaying trend. The concept of an effective temperature has also been introduced below to interpret the measured EPL.

Following the above analysis, $T_{eff}$ at different biases for EL can be extracted by linearly fitting the reduced spectra to the Boltzmann form. With the $T_{eff}$, we can then infer back to the underlying LSP contribution $\rho(\omega)$ of the junction from emission spectrum. As shown in Fig. 2C, $\rho(\omega)$ obtained from spectra at different voltages collapse to a single curve, confirming that the plasmonic resonances are an intrinsic property of the specific junction, independent of external stimuli.

We then proceed to analyze the measured EPL results from the same junction under optical illumination. By dividing the EPL spectra (the dashed red lines in Fig. 2D excluding the SERS contribution) by the inferred $\rho(\omega)$ from Fig. 2C, we can test for and obtain a similar exponential dependence. The resultant reduced spectra plotted on log scale (excluding the anti-Stokes SERS of organic contaminants)(Fig. 2E) again show clear linear frequency dependence, similar to that obtained in EL (Fig. 2B), indicating that an effective temperature can still be used as a parameter to describe the EPL of tunnel junctions.[26]

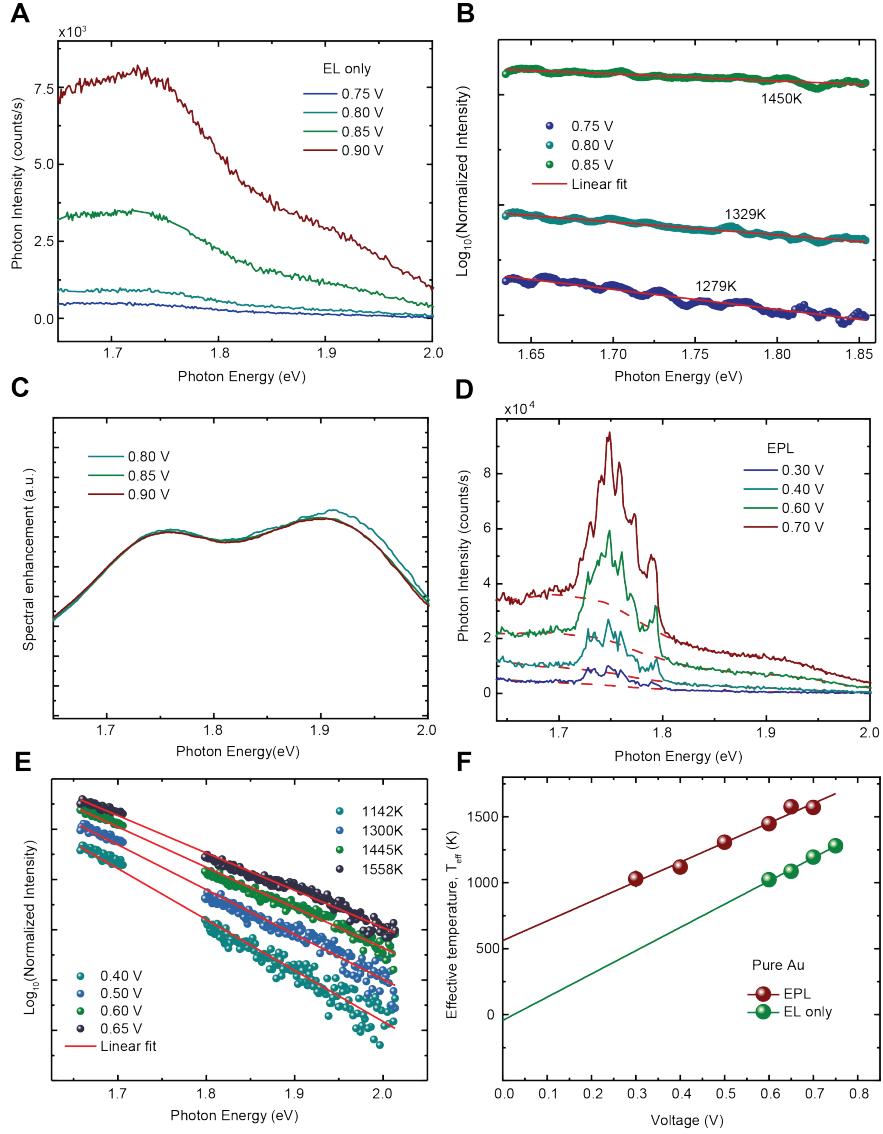

**Figure. 2.** Measurement and analysis of voltage-dependent upconversion light emission. (**A**) Spectral emission for an electrically driven Au junction (EL only) at different biases. (**B**) Normalization analysis of the spectra in **A** (spectra at 0.75V, 0.80V and 0.85V are normalized by the spectrum at 0.90 V). The linear decay of the reduced spectra with energy, plotted on log scale, is fitted with a Boltzmann distribution, $e^{-\hbar\omega/k_B T_{eff}}$, where $T_{eff}$ is the effective temperature of hot carriers. (**C**) Extracted voltage-independent plasmonic function of the junction from **B**. (**D**) Measured EPL for the same junction at different biases. The incident laser power is 0.46 mW. The dashed red lines correspond to pure EPL excluding the contaminant anti-Stokes SERS contribution. (**E**) Normalization analysis of EPL spectra, by dividing the spectra in **D** by the plasmonic function in **C**. Red lines represent the linear fit with Boltzmann distribution. The energy range with significant SERS was excluded. (**F**) Inferred $T_{eff}$ for EPL (red) and EL (green) vs. the applied voltage. The lowest measurement voltage (0.6 V for EL and 0.3 V for EPL) is limited by the noise level of the CCD spectrometer. Error bar represents the standard deviation of linear fit in **E.**

In Fig. 2F, we plot the relation between the applied bias and the extracted $T_{eff}$ of both EL and EPL in the Au junction. We can see that in both EL and EPL, $T_{eff}$ found from the Boltzmann analysis increases linearly with the applied bias. Moreover, a significantly higher $T_{eff}$ is found in the presence of optical pumping (EPL) than the $T_{eff}$ in the EL-only case. Linearly extrapolating the EPL-inferred $T_{eff}$ data down to zero bias (EPL → PL), we find a non-zero $T_{eff}$ at this limit (~500 K in Fig. 2F). In contrast, extrapolating the EL-inferred $T_{eff}$ data toward zero bias yields $T_{eff}$ ($V$→0) close to 0. The Boltzmann factor and the large difference in $T_{eff}$ between EPL and EL are the reason for the giant synergistic increase in upconversion emission in this hot carrier picture. By contrast, control experiments using a thin (~1nm) Cr adhesion layer as a damping medium for the plasmonic response[37] shows a much smaller $T_{eff}$ difference between EPL and EL and a less dramatic increase (<~10×; see Supporting Information Sec. 7 for the experimental results of Au/Cr junctions), suggesting the dramatic increase of upconversion emission in EPL is also closely related to the strength of plasmonic resonance of the materials.

We now focus on the dependence of the EPL on the optical power in tuning the increased emission and influencing the effective temperature. Figure 3A shows the results of the emission spectra for an Au junction under fixed bias $V$ = 0.6 V and varying optical power. As expected, higher excitation power induces stronger light emission. Moreover, $x(V$ = 0.6 V, $P)$ increases with $P$, reaching over 700 for relatively small laser power ( <1 mW), as shown in Fig. 3B.

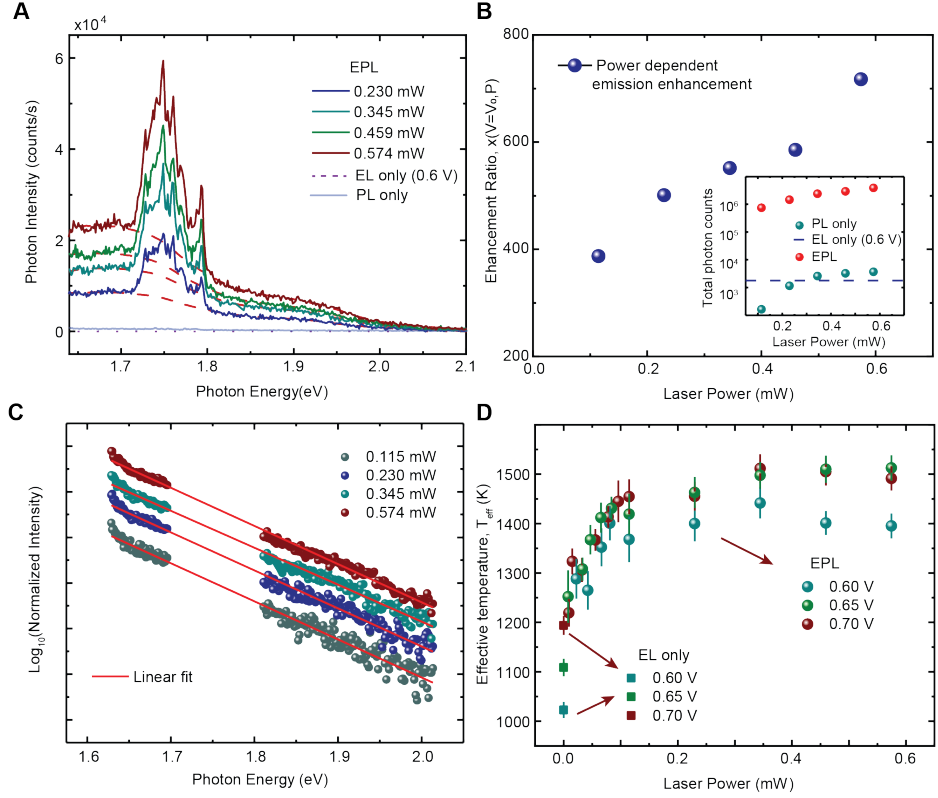

**Figure 3.** Measurement and analysis of power-dependent upconversion light emission in Au junctions. (**A**) Spectral light emission for EL, PL, and EPL at different laser power. Dashed red lines are the pure EPL spectra excluding contaminant anti-Stokes SERS contributions. (**B**) Measured enhancement ratio vs. power for light emission recorded in **A** with applied bias $V_0 = 0.6$V. Inset shows the calculated total photon counts for EL, PL, and EPL, respectively. The photon count of EL is indicated by the dash line. (**C**) Normalization analysis of the power-dependent light emission spectra in **A**. Red lines represent the linear fit with a Boltzmann distribution of hot carriers from which the effective temperature of hot carriers can be extracted. The anti-Stoke SERS portion has been removed in this analysis. (**D**) Inferred power-dependent effective temperature and comparison with the effective temperature obtained for EL under the same voltage bias. Error bars represent the standard deviation of the linear fits in **C**.

We performed the normalization analysis to extract the $T_{eff}$ for different $P$, with $\rho(\omega)$ obtained by the same analysis for EL spectrum described above. It can be seen from Fig. 3C that the logarithmic normalized intensity again exhibits excellent linearity with photon energy. Plotting all extracted $T_{eff}$ at different biases and power (Fig. 3D), a non-linear power dependence of $T_{eff}$ is revealed, in strong contrast to the linear relation between $T_{eff}$ and applied bias (Fig. 2E and 2F). Moreover, we find that $T_{eff}$ increases with laser power rapidly and saturates to a voltage-dependent value. These observations, combined with the measured increase effects shown in Fig. 3B and Fig. 1D, suggest a non-trivial role of combined optical and electrical excitations.

**Discussion**

We proceed to consider the physical mechanisms behind the observed light emission increase. Here we discuss two models with distinct microscopic electronic dynamics that can yield the observed exponential dependence on emission energy.

Firstly, following our previous work for EL[24], we extend the microscopic model based on hot carrier dynamics to elucidate the mechanism of the observed increase in upconversion emission. We found that the resultant EL spectra can be modelled by

$$U(\omega, V) \propto I^\alpha \rho(\omega)\, e^{-\hbar\omega/k_B T_{eff}} \quad (2)$$

where $\rho(\omega)$ is the LSP contribution, $I$ is the tunneling current, $\alpha$ is experimentally extracted value that was found always greater than 1 and indicates the nonlinear current-dependence of the above-threshold emission. Note that the hot carriers are not confined only to the drain electrode, but are generated throughout the region where the damping of LSPs occurs[38].

A more detailed treatment[39] looks at energy dissipation and transport of hot carriers within the electronic system as a consequence of the electronic viscosity at high current density, obtaining the same voltage dependence of the effective temperature ($T_{eff} \propto V$).(see Supporting Information Sec. 8 for the detailed derivation of the model and calculation for $\gamma_{e-e}$) The effective temperature of hot carriers is given by

$$T_{eff} = \sqrt{(\gamma_{e-e}V)^2 + T_0^2} \quad (3)$$

where $\gamma_{e-e}$ is related to the effective electronic viscosity and $T_0$ is the temperature of background electrons in the electrode. Assuming $T_0 \ll T_{eff}$, valid in the absence of illumination, we will obtain the linear voltage-dependent effective temperature $T_{eff} \approx \gamma_{e-e}V$. In previous work[39], calculation for a Au quantum contact predicts $\gamma_{e-e} \approx 65$ K/V, but neglects any possible contribution of electronic coupling to LSPs that would significantly amplify the electronic friction. Using the plasmonic properties of Au[40] and applying the formula to calculate the shear viscosity of an electron fluid[41], we estimate a revised value of $\gamma_{e-e} \approx 1495$ K/V, which agrees very well with our measurements in Fig. 2F (~1600 K/V for EL).

In this picture, under simultaneous optical and electrical excitation, there exists an optically driven additional incoherent temperature $T_0$ in the electrode. Therefore, $T_{eff}$ will show the interaction between optically and electrically generated hot carriers. We follow the approach of Liu et al.[42] to model the heating process due to optical excitation of LSPs which subsequently decay into hot carriers nonradiatively. Voltage and power dependence for modeled $T_{eff}$ under different excitation condition are plotted in Fig. 4A and 4B.

As shown in Fig. 4A, $T_{eff}$ under optical excitations shows a non-zero value in the zero-bias limit, consistent with our measurement (Fig. 2F). Moreover, the observed nonlinear behavior of power-dependent $T_{eff}$ and the saturation at high power with voltage dependent saturation value are well reproduced in the model (Fig. 4B).

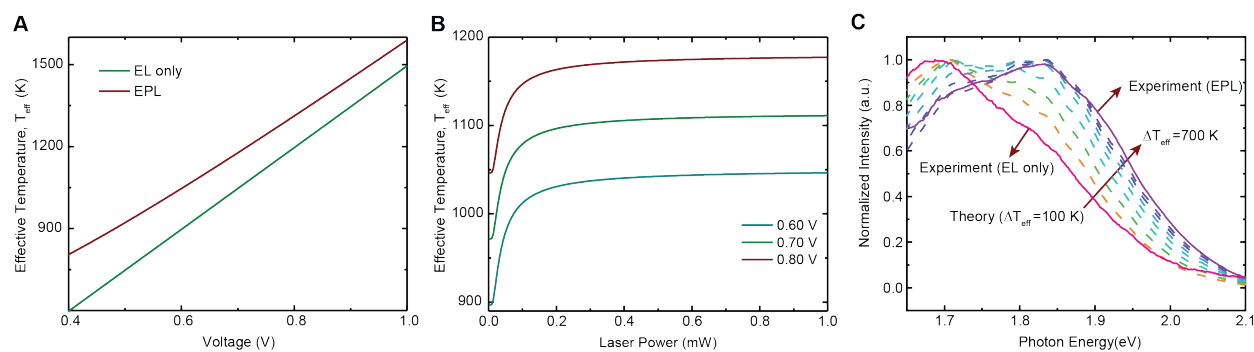

**Figure 4.** Theoretical modelling of upconversion light emission. (**A**) Calculated voltage dependence of the effective temperature of hot carriers in EL (green) and EPL (red) cases. (**B**) Calculated optical power dependence of the effective temperature of hot carriers in EPL at different bias voltages. (**C**) Starting from the experimentally measured EL spectrum (red), the inferred plasmon spectral shape is combined with Boltzmann factors of increasing $T_{\text{eff}}$ (calculated dashed curves). The calculated spectra blue shift and match the experimental EPL spectrum (purple solid curve) with a sufficiently large increase in effective temperature.

While not meant to be a detailed theoretical treatment of this complex system, this model can also quantitatively predict the upconversion emission spectra, as shown in Fig. 4C. Given the measured EL and EPL (peak magnitude normalized; the full spectra without normalization and the comparison with theory are shown in Fig. S7) of a Au junction (the solid lines in Fig. 4C), it can be seen that by taking the EL-inferred spectral shape ($\rho(\omega)$) and manually increasing $T_{eff}$ in Eq. (2), the calculated spectrum exhibits a clear blue shift due to the variation of the Boltzmann factor, and eventually reaches a near-perfect match with the normalized, measured EPL, further suggesting that both EL and EPL can originate from the same hot carrier effect. Moreover, we can calculate the EPL spectrum by replacing the effective temperature in the EL spectrum with the effective temperature under combined external stimuli and combining a pre-factor extracted from the fitting for the EPL spectrum. The calculated EPL spectrum agrees very well with the measured one, as shown in Fig. S7.

Within the hot carrier model, the generalized formula for EPL/EL/PL we derived enables the numerical estimation of the enhancement ratio (see Eqs. S15 and S16). The key to the synergistic effect lies in the optically induced incoherent temperature $T_0$, which induces a slower exponential decay with energy at the same bias and thus enabling a more prominent contribution of the LSP resonance, both in shape and amplitude. It can be seen that the cooperative effect will be most pronounced at a (device dependent) moderate bias, where the saturated $T_0$ value gives the largest difference in the exponential factor before and after optical illumination, explaining why the enhancement ratio decreases with bias in Fig. 1D. Such an optimum bias represents the energy scale where the electrical excitation is comparable to the optical pumping. Although the devices in this work do not possess a high energy conversion efficiency (~$10^{-6}$ or below), we note that further improvement can be readily foreseen by using better plasmonic materials, e.g., single-crystalline Au or Ag[13], and employing tunneling devices with high stability under high power.

In addition to the hot carrier recombination mechanism discussed above, anti-Stokes electronic Raman scattering is also a relevant physical mechanism in illuminated, biased junctions and could result in a similar light emission spectrum. In the anti-Stokes electronic Raman process, an electron above the Fermi level is excited by an incident photon into an intermediate virtual state, and subsequently recombines with a hole at a different energy and momentum, resulting in a blue shifted photon scattered into the far field[30,31]. The electronic crystal momentum change during this process, $\Delta k$, is constrained by the energy dispersion in the metal, which greatly suppresses this process in the metallic bulk. However, this constraint is relaxed near the surface of nanostructures, where in real space the energy-dependent overlap of the electronic wavefunctions results in a transition rate that includes a factor which decays exponentially with the Raman shift[31]

$$\Gamma(\epsilon) \propto \int f(E)[1 - f(E + \epsilon)] e^{-|2\epsilon/\Delta|} dE \qquad (4)$$

where $f(E)$ is the Fermi–Dirac distribution for an electron with energy $E$. and $\Delta$ is the exponential energy decay that may be estimated from a jellium model. The scattered photon then couples to the localized surface plasmon and results in far field emission. In the absence of a voltage bias, the anti-Stokes emission is limited by the finite carrier temperature of the Fermi-Dirac distributions.

A voltage bias can strongly affect anti-Stokes emission in the electronic Raman process by energetically allowing Raman scattering from electron states at the higher Fermi level electrode to hole states at the lower Fermi level electrode. This can lead to anti-Stokes emission within the scale of the bias window $eV$, but exponentially suppressed with the Raman shift on the scale of $\Delta$.

In this picture, the exponentially decayed normalization spectrum as a function of photon energy (Fig. 2E and 3C) can be considered using Eq. (4) with a bias-dependent allowed energy.

From the pure electroluminescence data, the hot carrier emission mechanism clearly takes place in these junctions. Similarly, electronic anti-Stokes Raman scattering has been demonstrated in other systems.[31] Determining the relative contributions of the two mechanisms in the present experiments is challenging, since both predict increased emission under combined excitation and the exponential decay of above-threshold emission as a function of energy. The bias and optical power dependence of the experimental data can provide insights for the different microscopic processes described by these two models. While not conclusive, the observation (Fig. 4C) that the electroluminescence spectrum and a modeled change in effective electronic temperature can give the combined EPL spectrum is suggestive. Conclusively discerning the respective contributions from these two distinctive mechanisms from CW measurements alone is very challenging and beyond the scope of this work. While the electronic Raman process is prompt[30], in the hot carrier recombination process the carrier scattering time is much slower (hundreds of femtoseconds)[26]. This difference may allow ultrafast measurements to disentangle the respective contributions of hot carrier dynamics and bias-assisted anti-Stokes electronic Raman scattering.

Our work provides insight into non-trivial light matter interactions at the nanoscale, demonstrating that the combination of abundant electronic tunneling and interactions with local plasmonic excitations can generate energetic photons in multiple ways. The experimental strategy employed here and the observation of increased emission under combined electronic and optical excitation opens numerous opportunities in nanophotonic and nanoplasmonic applications such as optical modulation of deep sub-wavelength plasmonic photon sources and hot-carrier photochemistry.[3,5,43].

## ASSOCIATED CONTENT

**Supporting Information**.

The following file is available free of charge on the ACS Publication website. Nanofabrication and Sample Preparation; Electromigration Break Junction Protocol; Experimental Setup and Measurement Procedures; Influence of anti-Stokes SERS Signal on Measured Light Emission Spectra; Analysis on the Stokes side spectrum; Data Processing and Normalization Analysis; Control experiments on Au/Cr junctions; Theoretical Model of Hot Carrier Dynamics Induced Light Emission.

## AUTHOR INFORMATION

**Corresponding Author**

*Email: natelson@rice.edu

**Author Contributions**

D.N and L.C. designed the experiment. L.C. and Y.Z. fabricated the devices, conducted the experiment, and modelled the data. M. D. modeled viscous electronic heating in the presence of plasmons, and P.N. theoretically modelled the optically driven hot carrier system. All authors wrote the manuscript and have given approval to the final version of the manuscript.**Notes**

The authors declare no competing financial interest.

**Funding Sources**17

J. Evans Attwell Welch Fellowship. Robert A. Welch Foundation award C-1222. AFOSR award No. FA 9550-15-1-0022. NSF ECCS-1704625 and Robert A. Welch Foundation award C-1636.

## ACKNOWLEDGMENT

L.C. acknowledges funding support from J. Evans Attwell Welch Fellowship, the Smalley Curl Institute at Rice University, and the support of College of Engineering and Applied Science at University of Colorado at Boulder. D.N. and Y.Z. acknowledge Robert A. Welch Foundation award C-1636. P.N. acknowledges Robert A. Welch Foundation award C-1222 and AFOSR award No. FA 9550-15-1-0022. D. N. acknowledges support from NSF ECCS-1704625.

## ABBREVIATIONS

LSP, localized surface plasmon.

EL, electroluminescence.

PL, photoluminescence.

EPL, electro-photo-luminescence.

**For Table of Contents Only**

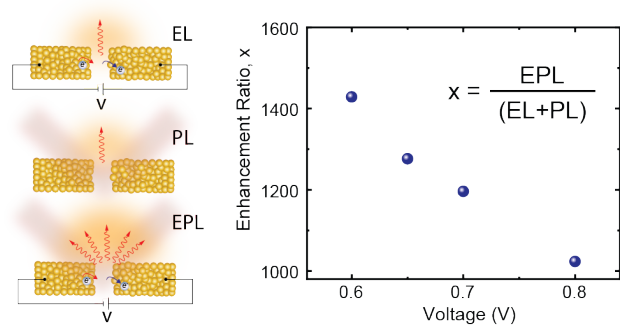

# Supporting Information

# Thousand-fold Increase in Plasmonic Light Emission via Combined Electronic and Optical Excitations

*Longji Cui[1,2,3,†], Yunxuan Zhu[1,†], Peter Nordlander[1,4,5], Massimiliano Di Ventra[6], Douglas Natelson[1,4,5,*]*

[1]Department of Physics and Astronomy and Smalley-Curl Institute, Rice University, Houston, TX 77005, United States

[2]Paul M. Rady Department of Mechanical Engineering, University of Colorado, Boulder, CO 80309, United States

[3]Materials Science and Engineering Program, University of Colorado, Boulder, CO 80309, USA.

[4]Department of Electrical and Computer Engineering, Rice University, Houston, TX 77005, United States

[5]Department of Materials Science and Nanoengineering, Rice University, Houston, TX 77005, United States

[6]Department of Physics, University of California San Diego, La Jolla, CA 92093, United States

[†]These authors contributed equally to this work.

[*]Corresponding author: Douglas Natelson (natelson@rice.edu.)

**Supplementary Information Text**

**1. Nanofabrication and Sample Preparation**

We fabricated the array of nanowire devices on a 500 μm thick Si wafer topped with a 200 nm-thick wet thermal SiO$_2$ layer. Large electrode pads (50 nm Au thick with a 5 nm Ti adhesion layer) are deposited on the wafer using shadow mask evaporation. Lithographically defined nanowires together with the bow-tie shaped fan-outs were patterned by e-beam lithography. The nanowire geometry (100 nm wide, 600 nm long and 18 nm thick) had been optimized through numerical simulations to enhance the plasmonic response of nanostructures at 785 nm (the applied continuous wave (CW) wavelength). Two types of devices were prepared using e-beam lithography, pure Au and Au/Cr junctions. A thin layer (~1 nm) of Cr was chosen in Au/Cr junctions as both an adhesion layer and as an effective plasmonic damping medium[1] for the control measurements to establish the role of LSPs in the observed light emission. Bilayer e-beam resist (PMMA 495/950) was used for pure Au devices in order to obtain high lift-off quality, while single layer e-beam resist (PMMA 495) was employed for devices with Cr adhesion layer. 18 nm thick Au (1 nm Cr if adhesion layer is needed) was then deposited using e-beam evaporation. Devices were cleaned using multiple cycles of oxygen plasma after wire-bonding and transferred immediately into the vacuum cryostat (Montana Instruments) for optical and transport experiments. We note that minute amount of organic contaminants from the fabrication and sample preparation processes cannot be entirely avoided on the nanowire sample, which, via plasmonic enhancement in close proximity to the tunnel junctions, leads to the observed anti-Stokes SERS features on the light emission spectra.

**2. Electromigration Break Junction Protocol**

Tunnel junctions were created on the metal nanowires using electromigration break junction technique[2]. Our previous work[3] shows that in order to observe upconverted photons from electrically driven tunnel junctions, a relatively high tunneling current (>10 μA, corresponding to an electrical conductance of ~0.1 $G_0$ under 1-2 V of voltage bias) is required. In order to form such tunnel junctions with high yield, the electromigration process was first implemented at 80 K by applying cycles of sweep voltage (rate of 10 mV/s) to thin the nanowire, indicated by a small sudden drop in the electric current exceeding a pre-set value (0.4% of the initial resistance in the beginning of each cycle). Once the resistance of the nanowire increased to ~500 Ω, indicating a sufficiently thinned nanowire, the substrate temperature was lowered to 30 K. Further electromigration was then implemented until the resistance suddenly increases above 12.9 kΩ (1 $G_0$), resulting in a tunnel junction with a desirable conductance for light emission measurements. To maintain high stability of the junctions, all measurements were performed at 30 K (substrate temperature). The *I-V* characteristics of the electromigrated tunnel junctions were monitored continuously throughout the measurements to ensure there was no discernable conductance change of the junction. To ensure the reproducibility of the measurement results, the light emission spectra were repeatedly measured at different electrical biases and optical powers for each junction studied.

## 3. Experimental Setup and Measurement Procedures

Following electromigration of the Au nanowire, we applied a linearly polarized 785 nm CW laser (corresponding to the resonance wavelength of the transverse plasmon mode of Au nanowire) as the excitation light source and focused the laser beam through a high NA objective (Nikon 50×, NA 0.7) with a diameter of 1.8 *μ*m on the junction. A home-built Raman setup was used to align the free space optics through which the emitted photons were filtered (BNF 785) and the spectrum was collected using an optical spectrometer (Horiba iHR 320/Synapse CCD). Electrical transport through the junction can be simultaneously measured by a current pre-amplifier (SRS 570) in the presence of electrical bias. To maintain the high stability of the tunnel junction (avoiding small configuration change of the junction), we applied the laser power of <1 mW and electrical bias of <1.5 V. All light emission spectra shown in the main text were taken by averaging five consecutively recorded spectra (each with 5s acquisition time). Due to ultraminiaturized junction size (on the sub-nanometer scale) compared to the size of focused laser beam, we scanned the laser beam, with the position controlled by a nanopositioner (ANC 300 Piezo Controller), to identify the exact spot on the tunnel junction with the maximal light emission enhancement. The results were compared with the spatial mapping of the open-circuit photo-voltage (OCPV) signal using OCPV scanning microscopy technique described below to confirm the key role of photogenerated hot carriers.

**4. Influence of anti-Stokes SERS Signal on Measured Light Emission Spectra**

Plasmonic tunnel junctions studied in this work are associated with large radiative field enhancement effect due to the LSPs confined in the tunneling gap. Past studies[4,5] have shown that SERS features even on the single-molecule scale can be identified in such plasmonic nanostructures. Our device nanofabrication and sample handling process may inevitably introduce residual organic molecules deposited on the nanowire, potentially near the tunnel junction, leading to measurable SERS signal superimposed on the EPL spectra. Hence, our data analysis has excluded the observed anti-Stokes surface-enhanced Raman spectroscopy (SERS) features shown on the recorded spectra (peaks of Raman-active phonon modes), which originate from residual organic molecules (e.g., hydrocarbons from nanofabrication and sample preparation, corresponding to Raman shifts of ~1000-1800 cm$^{-1}$) in proximity to the tunneling gap. Due to device-to-device variation it is common to have varying amounts of residual organic molecules in different tunneling gaps, leading to distinct molecule Raman peak intensity for individual junctions (e.g., very little for the device in Fig. 1; considerably more for the device in Fig. 2). In Fig. S1, we have shown a EPL spectrum with the assigned SERS peaks corresponding to the specific hydrocarbon signatures[6,7]. In our data analysis, we have excluded the anti-Stokes SERS peaks by extracting the broadband trendline (the dashed line in Fig. S2) on the emission spectra. As shown in Fig. 2 to 3 in the main text, the anti-Stokes regions have been removed when performing normalization analysis and inferring the effective temperature on the light emission spectra. It can be seen in Fig. 2E and Fig. 3C, that the effective temperatures extracted in the energy range below 1.7 eV and that above 1.8 eV showed very nice consistency.

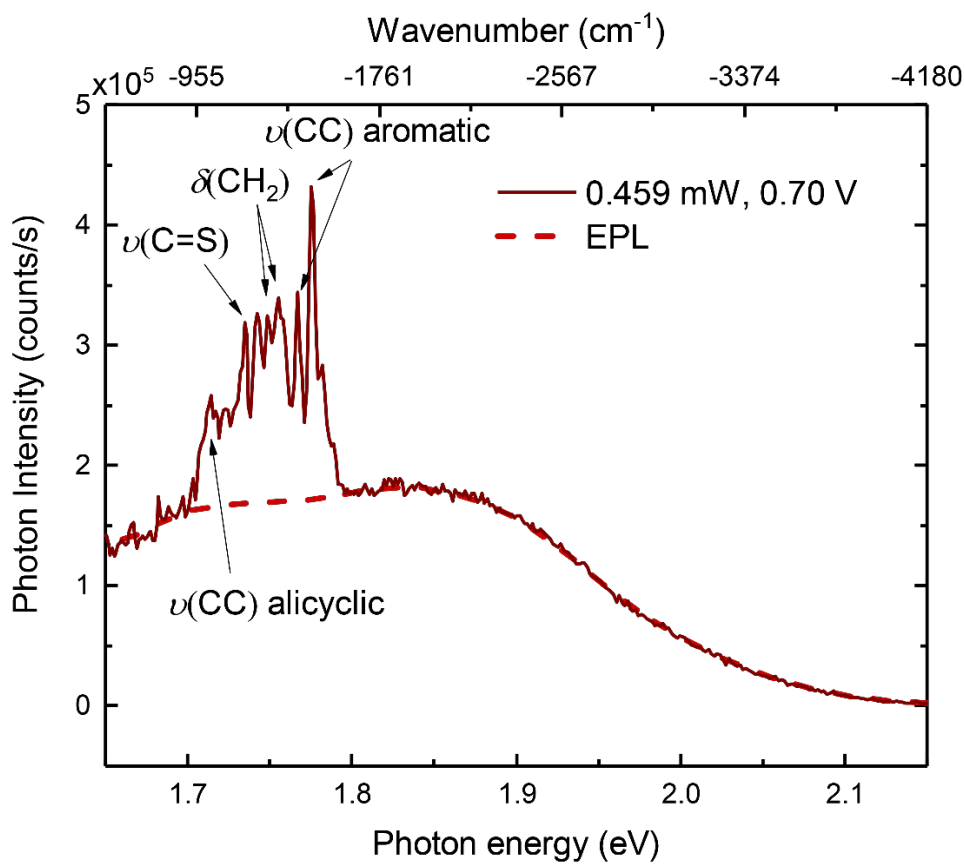

**Figure S1.** Anti-Stokes SERS features on the measured light emission spectrum from another tunnel junction different those shown in the main text. The arrows in the inset mark the different vibrational modes' position for residual organic molecules. Dashed red lines are the extracted pure EPL spectra with anti-Stokes SERS contributions excluded, based on the pure EL spectra for this device.

## 5. Analysis on the Stokes side spectrum

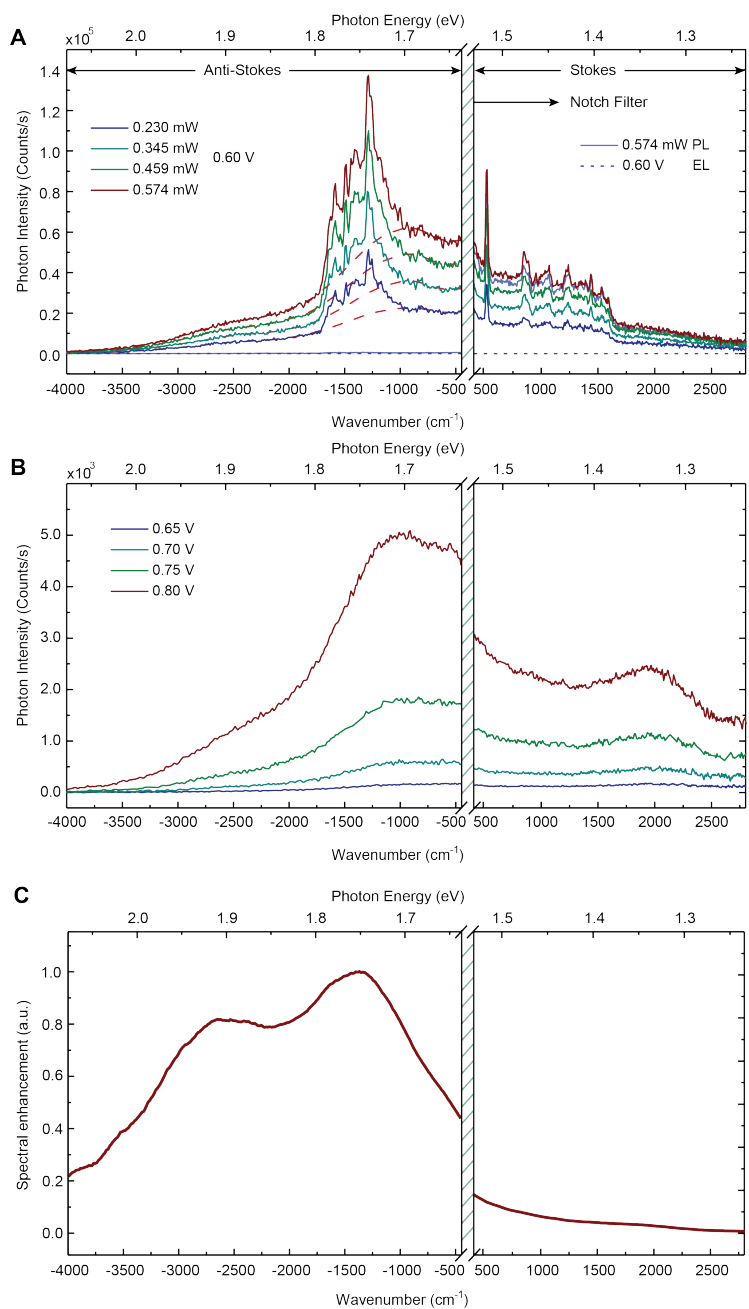

**Figure S2. Full light emission spectrum and analysis results.** (**A**) Measured EPL spectra as a function of the incident laser power on the anti-Stokes side combined with the simultaneously measured data on the Stokes side, with the PL and EL spectra under respective bias and power also included. The sharp peak on the Stokes side is the silicon peak from the substrate at 520cm$^{-1}$. The region from -400 cm$^{-1}$ to 400 cm$^{-1}$ (shaded area, notch filter applied) is removed. (**B**) Full EL spectra for the same tunnel junction at different biases. (**C**) Extracted plasmon resonance after normalization analysis on both sides. Note the significant resonance peak occurs around 1.75 $eV$, which leads to the pronounced emission and vibrational mode pumping at the anti-Stokes region in **A**.

One may question whether the giant enhancement effect in the anti-Stokes light emission region could be a natural consequence of the extremely inefficient anti-Stokes emission, which often only account for a tiny fraction of the total light emission; that is, the enhancement could be a result of dividing by a small denominator. To address this concern, we performed measurements of both the Stokes and anti-Stokes emission. In these devices, the anti-Stokes emission is comparable and even more pronounced compared to the counterpart Stokes emission. Figure S2 shows the full light emission spectrum (both Stokes and anti-Stoke emission) of a Au tunnel junction at different optical powers. It can be clearly seen from Fig. S2A that the anti-Stokes emission is comparable or even stronger in magnitude to the Stokes counterpart, both for the molecular vibrational modes and the broadband hot carrier light emission (which is the focus of our work). The EL spectra (Fig. S2B) under different voltages bias also exhibit a similar behavior. To understand this phenomena, we performed a normalization analysis, described thoroughly in our previous work[3] and in the analysis of Fig. 2 of the main text, to extract the LSP resonances of the tunnel junction. As shown in Fig. S2C, the spectral enhancement due to the LSPs is significantly higher on the anti-Stokes side (with peak resonances at ~1.7-1.8 eV) than that on the Stokes side. This explains why the anti-Stokes emission in our measurements is much more intense and the enhancement effect in upconversion photon emission which is from the plasmon-induced hot-carrier process is so dramatic. It differs from typical SERS analyses which often assume a weakly energy-dependent plasmonic response of the substrate.

Furthermore, we can also ask whether tunneling of photoexcited carriers followed by inelastic plasmon excitation could be responsible for the observed broadband emission. However, such a mechanism can be largely ruled out given the fact that the photo-generated current accounts for less than 1% of the total tunneling current under simultaneous electrical and optical excitations (see Fig. S3 for an example measurement result). Such a tiny increase in the total tunneling current is unlikely to be responsible for the orders-of-magnitude above-threshold light emission enhancement.

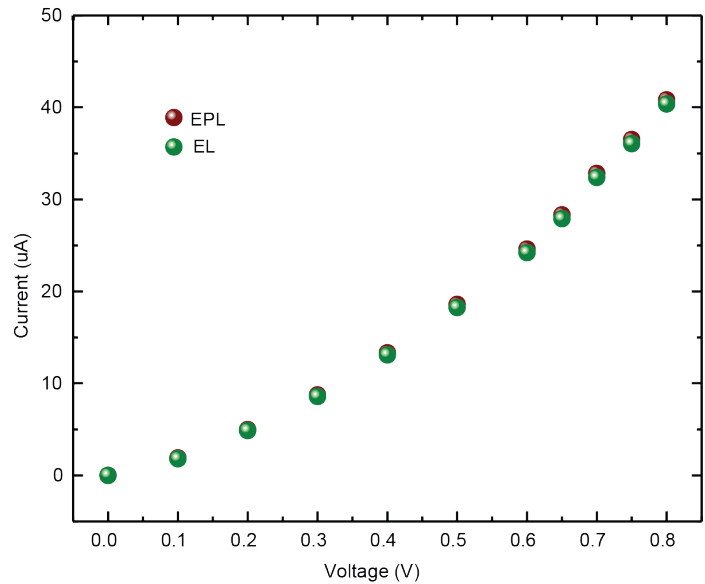

**Figure S3**. An example of $I-V$ characteristics for both EL and EPL for the tunnel junction in Fig. 1. Note that photon-generated current (difference in the total currents under EL and EPL excitations) accounts for a very small portion (<1% below 0.8V) of the total tunnelling current.

### 6. Data Processing and Normalization Analysis

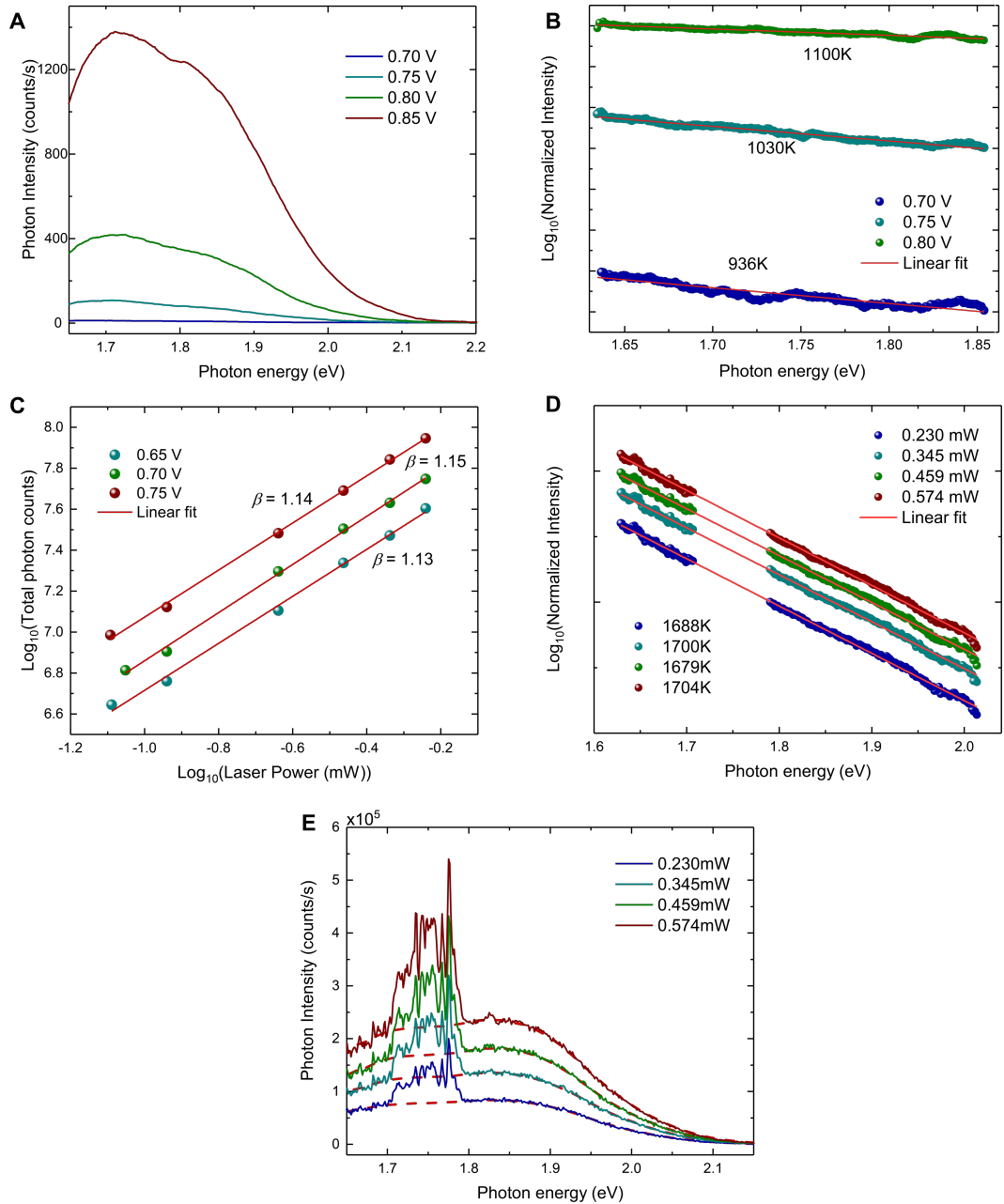

**Figure S4**. Additional example of data processing and normalization process. (**A**) EL spectra light for an electrically driven Au junction at different biases. (**B**) Normalization analysis of the spectra in **A** (the emission spectra at 0.70V, 0.75V, and 0.80V are normalized by the spectrum at 0.85 V). The linear decay of the reduced spectra with energy, plotted on log scale, is fitted with a Boltzmann distribution. (**C**) Measured total photon counts as a function of the applied optical power in EPL. A power law fitting is plotted to extract the exponent β. (**D**) Normalization analysis of EPL spectra, by dividing the measured spectra in **E** by the plasmonic function. (**E**) Measured EPL spectra for the same tunnel junction, with the dashed red lines corresponding to pure EPL excluding the contaminant anti-Stokes SERS contribution.

We developed a new normalization analysis method based on our previous work[3] to process the measured light emission spectrum by separating out the plasmonic modes that only depend on the geometric details of a tunnel junction. Here we provide an additional example (Fig. S4) to describe the detailed procedures in implementing this analysis. In our previous work[3] on EL from electrically driven tunnel junctions, we showed that the light emission spectrum, $U^{EL}(\omega)$, can be approximated by

$$U^{EL}(\omega) \propto I^\alpha \rho(\omega) \hbar \omega e^{-\frac{\hbar\omega}{k_B T_{eff}^{EL}}} \qquad (S.1)$$

where $I$ is the tunneling current, $\alpha = 1.2$ indicates the current dependence of EL from our previous statistical analysis on the pure EL, $\rho(\omega)$ is the local photon density of states modified by the plasmonic resonance of the junction, and $T_{eff}^{EL}$ is the effective temperature of electrically generated hot carriers for which we found $T_{eff}^{EL} \propto V$.

Given the measured EL spectra (Fig. S4A, as well as Fig. 2A in the main text), we first divide the spectrum at different bias with their respective current $I^\alpha$ and normalized them after dividing the current (i.e., 0.70 V, 0.75 V, and 0.80V) with reference to the spectrum at the highest bias (i.e., 0.85 V). By doing so all the terms in Eq. (S. 1) before the exponential factor are canceled out, leaving only the Boltzmann like factor, which is characterized by the effective temperature difference with the reference curve (highest bias in this case). This leads to the normalized spectrum at each bias as shown in Fig. S4B. When plotting on log scale (Fig. S4B and Fig. 2B), the linear dependence of the normalized spectra on photon energy strongly suggests a universal underlying physical process (i.e., the Boltzmann distribution of hot carriers, $e^{-\hbar\omega/k_B T_{eff}^{EL}}$) in generating the observed EL. $T_{eff}^{EL}$ was then extracted from the normalized spectra by fitting with the Boltzmann factor. For convenience, here the spectrum measured under the highest voltage is chosen for the normalization reference. In practice, choice of the normalization spectrum only weakly affects the values for the extracted effective temperatures. For example, 0.70 V, 0.75 V, and 0.80V can all be chosen as the line for normalization, while the extracted effective temperatures at each bias for different normalization reference are within five percent of variance. Such deviation is mainly due to the different noise levels at different biases affecting the fitting slope after normalization. Subsequent to the normalization analysis, we applied Eq. (S. 1) to extract $\rho(\omega)$ and confirmed that $\rho(\omega)$, as an intrinsic property pertaining to a specific tunnel junction.

In contrast to EL spectra, EPL spectra is expected to scale with both the tunneling current and applied optical power ($P$). To study the power dependence of EPL, we first performed measurements by fixing the applied bias and calculate the total upconverted photon counts as a function of the varied power. As shown in Fig. S4C, it can be seen that there exists an excellent linearlity between the total photon counts and the applied power, which suggests

$$\int U^{EPL}(\omega) \cdot d\omega \propto P^\beta \tag{S. 2}$$

where the power law exponent $\beta$ is found to be very close to 1. We note that an analogous analysis has been performed in our previous work to extract the tunneling current dependence of EL.

Based on the above analysis and inspired by Eq. (S. 1) for EL, we then proposed an expression to approximate the observed EPL spectra

$$U^{EPL}(\omega) \propto I^\alpha P^\beta \rho(\omega) \hbar\omega e^{-\frac{\hbar\omega}{k_B T_{eff}^{EPL}}} \tag{S. 3}$$

We note here that Eq. (S. 3) only approximates EPL in the limit of high optical and low EL intensity knowing that light emission should reduce to Eq. (S. 1) in the zero optical power limit. To verify the validity of this model, we first normalized the measured spectra (Fig. S4D, as well as Fig. 2E in the main text) by dividing the EPL spectrum by $\rho(\omega)$ which we have inferred from the normalization analysis of EL spectrum given the fact that the same tunnel junction (and thus invariant $\rho(\omega)$) was measured for both EL and EPL. The reduced EPL spectra at different biases and optical power, plotted on log scale (Fig. S4D, Fig. 2E), demonstrate clear linear dependence with photon energy, strongly suggesting the Boltzmann factor ($e^{-\hbar\omega/k_B T_{eff}^{EPL}}$) still constitutes a valid description of the hot carrier system in Eq. (S. 3). Slightly different from the EL case, here the effective temperature for EPL can be obtained by directly fitting the reduced EPL spectrum to the Boltzmann factor, without the effective temperature for the reference curve presenting in the exponential term, since here $\rho(\omega)$ is straightforwardly removed from Eq. (S. 3), as can be seen in Fig. S4D, Fig. 2E and Fig. 3C.

By employing Eq. (S. 3) in conjunction with our theoretical model of effective temperature of hot carriers under electrical and optical excitation, our analysis can be applied to predict the light emission enhancement ratio and EPL spectrum for electrical and optical conditions and other plasmonic materials that are not attainable in our measurements (e.g., higher electrical and optical power, single crystalline Au or Ag, etc.). Fig. 4C in the main text shows the predicted EPL spectrum that can be perfectly obtained in theoretical analysis based on the knowledge of the plasmonic modes of the junction ($\rho(\omega)$).

**7. Control experiments on Au/Cr junctions**

We further test the relevance of plasmons to the upconversion light emission process by varying the plasmonic materials, performing measurements on Au/Cr tunnel junctions. The thin Cr layer damps the plasmonic response of Au significantly, reducing the rate of plasmon-based hot carrier generation compared to pure Au junctions. Figure S5 shows the results for a Au/Cr junction under different biases without (Fig. S5A) and with (Fig. S5B) incident optical power at P = 0.46 mW. The measured EPL spectrum only exhibits about 25% enhancement at the maximum ($1.65\,eV$) relative to the EL spectrum under the same bias, associated with a less dramatic enhancement effect (<~10×; plotted in Fig. S5D). After extracting the $T_{eff}$ by doing the same analysis in the pure Au junctions, it can be seen in Fig. S5C that while the qualitative linear relation between $T_{eff}$ and voltage remains much the same as Fig. 2F, the difference in $T_{eff}$ between EPL and EL for the Au/Cr junction is much smaller compared to pure Au junction. In high-bias limit, we obtain the same $T_{eff}$ for both EPL and EL and there is no enhancement observed (enhancement ratio reduces to unity). These findings further suggest the plasmonic resonance of the materials plays an important role in controlling the $T_{eff}$ of the excited hot carriers and thus determining the dramatic enhancement of upconversion light emission in EPL.

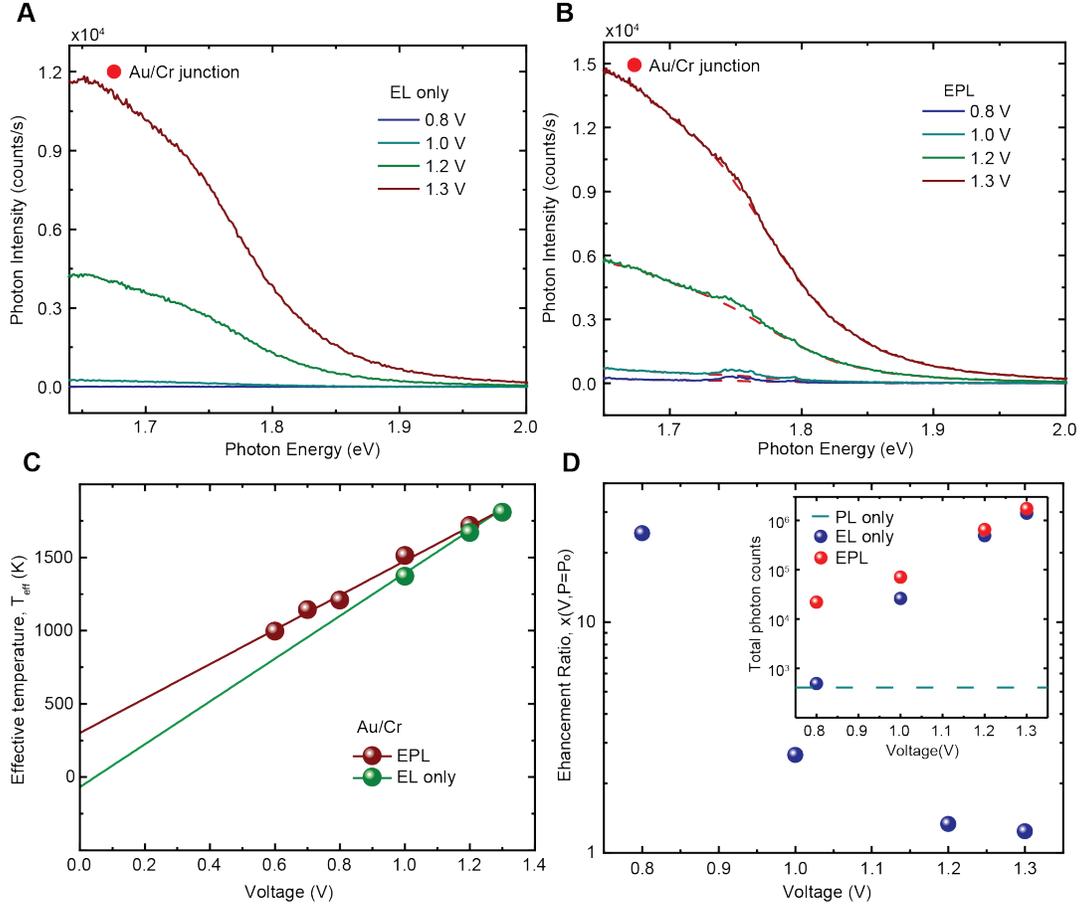

**Figure S5.** Control measurement of voltage-dependent upconversion light emission in Au/Cr tunnel junctions. (**A**) Spectral light emission intensity for an electrically driven Au/Cr tunnel junction (i.e., EL only) at different voltage biases. (**B**) Measured EPL spectra for the same tunnel junction. The applied laser power in EPL is 0.46 mW. (**C**) Inferred $T_{eff}$ for EPL (red) and EL (green) vs. the applied voltage. Error bar represents the standard deviation of linear fit in the fitting for the normalized EPL curve. (**D**) Calculated enhancement ratio at different bias with fixed incident laser power 0.46 mW applied laser power. Inset shows the calculated total photon counts for EL, PL, and EPL, respectively, at different biases.

## 8. Theoretical Model of Hot Carrier Dynamics Induced Light Emission

A complete, rigorous treatment of the biased junction under illumination is beyond the scope of the present work, requiring detailed and realistic modeling of the open quantum system including carrier transport, inelastic electron-electron and electron-phonon scattering, elastic scattering from disorder and surfaces, and interactions of carriers with the incident field. The key advance here over prior work[3] is the combination of both bias-driven and optically-driven carrier heating to produce an understanding of the observed nontrivial dependence of $T_{eff}$ on both bias and incident optical power (e.g., Fig. 2F), reproducing the nontrivial electrical bias and optical power dependence shown in Fig. 2F and Fig. 3D.

In the hot carrier model, the effective temperature $T_{eff}$ corresponds to the energy scale that quantifies the average energy of the excited hot carriers under quasi-equilibrium above the normal Fermi sea. In general, a driven, nonequilibrium system cannot be well described by a unique effective temperature[8]. Nevertheless, from our analysis above, the linear dependence of the logarithmic normalized intensity on photon energy shows that it is possible to define an effective Boltzmann factor for the carrier dynamics. As shown previously[3], measurements of EL in an electrically driven tunnel junctions show Boltzmann-like emission, with steady-state $T_{eff}$ linearly dependent on the applied bias $V$. Our theoretical model[3] suggests that this is a natural consequence of tunneling carriers exciting LSPs that then decay quickly into hot electrons and holes spread over the energy interval (-$eV$, $eV$) around the Fermi level ($\epsilon_F$). These hot carriers scatter off each other, while propagating and gradually losing energy to the lattice, leading to a steady-state Boltzmann tail of the electronic distribution with an average energy per carrier ($\bar{E}$), and hence $T_{eff}$ ($\bar{E} = k_B T_{eff}$) proportional to $V$.

Here, leveraging a more detailed treatment[9], hot carriers formed within the electronic system can be described in the language of electronic viscosity. Based on the energy dissipation and transport within the electronic system due to electronic viscosity, we will first derive the effective temperature of hot carriers, leading to a linear voltage dependence of effective temperature ($T_{eff} \propto V$) in the pure electroluminescence (EL) case. To elaborate, the extremely local character of tunneling in a junction results in a very high local current density and hence comparatively enhanced dissipative effects of shear viscosity in the electron fluid[9]. For a junction with transmittance $\tau$, the viscous dissipation is a fraction $\alpha$ of the total dissipated electrical power

$$P_{diss} = \frac{\alpha V^2}{R} = \alpha V^2 \tau G_0 \qquad (4)$$

where $G_0$ is the conductance quantum ($2e^2/h$) and $\alpha$ is a factor involving the material properties including the electronic viscosity. The dissipated power must be carried away by heat transport of the electronic system. Since for high-transmission junctions the most effective mechanism is elastic scattering, we can estimate this power dissipated according to the Landauer formula as

$$I_{th} \propto \int E\tau(E)[f_R(E, T_R) - f_L(E, T_L)] \, dE \qquad (S.5)$$

where $f_{L/R}$ is the Fermi-Dirac distribution of the left and right electrode at its own local temperature and chemical potential. Here we assume $T_L \equiv T_s$ and $T_R \equiv T_{eff}$ are the effective electronic temperatures, and energy-independent transmission $\tau(E) = \tau$ in the absence of electronic resonances. At low substrate temperature $T_s$ close to zero which is valid in the EL only scenario considering only cold background electrons in the electrodes as opposed to the excited hot carriers, the complete Fermi-Dirac integral for the second term in Eq. (S. 5) can be neglected and then Eq. (S. 5) is reduced to

$$I_{th} \propto \tau T_{eff}^2 = P_{diss} = \alpha V^2 \tau G_0 \qquad (S.6)$$

which yields the linear voltage-dependent effective temperature of hot carriers generating EL,

$$T_{eff} = \gamma_{e-e} V \qquad (S.7)$$

where $\gamma_{e-e}$ is related to the electronic viscosity and can be derived from microscopic theory based on a hydrodynamic approach[9],

$$\gamma_{e-e} = \left(\frac{G_0}{neA_c}\right)\left(\frac{d-1}{3d}\frac{\eta}{\gamma}\right)^{\frac{1}{2}}\left(3\left(\frac{r}{A_c}\right)^{1/2}\frac{L}{2}tan^{-1}\left(\left(\frac{r}{A_c}\right)^{1/2}\frac{L}{2}\right) + \frac{1-rL^2/4A_c}{(1+rL^2/4A_c)^2} - 1\right)^{1/2} \quad (S.8)$$

where $G_0 = 2e^2/h$ is the quantized unit of electrical conductance, $n$ is the electron density of the metal, $d = 3$ corresponds to the dimension of the system, $\eta$ is the shear viscosity of the electron liquid (~10$^{-7}$ Pa s) and its expression as a function of particle density is given in Eq. F8 in Di Ventra et al.[10], $\gamma = k_F^2 k_B^2 \lambda_e/9\hbar$ where $k_F$ is the Fermi momentum, $k_B$ is the Boltzmann constant and $\lambda_e$ is the electron mean free path. $A_c$ and $r$ are positive parameters that describes the geometric variance trend of the nano constriction with $A_c = 7.0Å^2$ [9] being the effective cross section for a quantum point contact. $L$ here is the characteristic length representing the changing of the cross section profile and satisfy the boundary condition that electron velocity quickly approaching the bulk value away from the junction: $A_c + rL^2/4 = mv_F G_0/ne$, where $v_F$ is the Fermi velocity[9]. The $\gamma_{e-e}$ value for a gold quantum point contact is estimated to be approximately $65K/V$ [9] which is more than an order of magnitude smaller than our experimental value from the extracted effective temperature. This large discrepancy is due to the omission of the important role of the Au plasmon in increasing the viscosity of the electron liquid. Increased viscosity coefficient for Au due to electron-plasmon friction can be estimated by treating electrons interacting with an effective 'plasmonic liquid' of appropriate parameters, by replacing the electron mean free path $\lambda_e$ with the effective Debye length $\lambda_F \approx 1Å$[11] which is the quantum analog of the Debye length for Au plasmon in the extreme vicinity of the tunnel junction where the LSP resonance is strong. This effective length leads to an increased mass density by a factor of 3.3, and an increased viscosity of $1.3 \times 10^{-6}$ Pa s, as estimated from the high-density limit of Eq. F.8[10]. Assuming all other geometrical factors in Eq. (S.8) to be the same as in D'Agosta et al.[9], we then estimate an increase of $\gamma_{e-e}$ by a factor of about 23 resulting in $\gamma_{e-e} = 23 \times 65 = 1495 K/V$. Note here that in deriving Eq. (S.7), we assumed the main mechanism for energy dissipation in the tunnel junction is elastic energy transport among the electrons. In the case of very low transmittance (much lower than in our experiment) energy dissipation via inelastic (non-electronic) channels becomes significant so that Eq. (S.7) may no longer hold.

We then focus on the effective temperature of optically generated hot carriers (pure photoluminescence (PL) case). In analogy with our previous experimental work[3], similar argument can be made for the formation of steady-state carrier distribution due to optical process. Hot carriers generated by the optically excited plasmons will scatter off each other many times, before finally thermalizing with the lattice via electron-electron and electron-phonon scattering. As the time spacing $\tau_p$ between successive arriving photons keeps decreasing (raising laser intensity) and reaches comparable or smaller values of carrier thermalization time with the lattice, a steady-state distribution of the hot carriers that thermally detaches from the lattice phonons can then forms as a consequence of persistent optical generation events preventing the complete thermalization of hot carriers. The continuously excited hot electrons and cold holes distribute symmetrically around Fermi level, hence for the following modeling we will deal with the part above the Fermi energy only.

To model the optically driven hot carrier temperature $T_0$, we proceed to analyze the relaxation dynamics of hot carriers that are generated by non-radiative decay of LSPs in the tunnel junction. Following an approach of Liu et al.[12], we define $T_0$ as the average energy content ($\bar{E}_{hc}$) of the hot carriers

$$k_B T_0 = \bar{E}_{hc}(P) = \langle \frac{\int d\varepsilon \cdot \varepsilon p_E(t,\varepsilon)}{\int d\varepsilon \cdot p_E(t,\varepsilon)} \rangle_{ss} \quad (S.9)$$

where $p_E(t,\varepsilon)$ is the population distribution of carriers at energy $\varepsilon$ and time $t$ and $\varepsilon$ is taken as the energy difference relative to the Fermi level $\varepsilon = E - E_F$.

Assume the temporal interval between consecutively excited LSPs (under CW laser illumination) is $t_p = \beta/P$, where $\beta = \hbar\omega/\sigma$, such that $\hbar\omega$ is the incident photon energy, $\sigma$ is the absorption cross-section of the metallic nanostructure, and $P$ is the incident intensity (the applied optical power per unit area of illumination). As shown previously in Liu *et al*[12], a sufficiently short $t_p$ compared to the the relaxation of a photogenerate hot carrier would allow a steady-state population of hot carriers. Specifically, at a particular energy level $\varepsilon = E - E_F$, assuming exponential hot carrier relaxation rate and that plasmon-excitation/carrier-creation events occur at equally spaced time intervals $t_p$ with each events creating $r(\varepsilon)$ number of hot carriers at $\varepsilon$ above Fermi level, then the steady state populations $p_{ss}(\varepsilon, P)$ at $\varepsilon$ can be found by summing infinite geometric series giving

$$p_{ss}(\varepsilon, P) = r(\varepsilon)(e^{-t_p/\tau_\varepsilon} + e^{-2t_p/\tau_\varepsilon} + e^{-3t_p/\tau_\varepsilon} + \cdots) = r(\varepsilon) \frac{exp\left(-\frac{\beta}{P\tau_\varepsilon}\right)}{1 - exp\left(-\frac{\beta}{P\tau_\varepsilon}\right)} \quad (S.10)$$

where $\tau_\varepsilon$ is the hot carrier lifetime at $\varepsilon$. According to Fermi liquid theory, $1/\tau_\varepsilon$ scales as $(E - E_F)^2$, giving $1/\tau_\varepsilon = \zeta\varepsilon^2$ with $\zeta$ being a proportional coefficient. Qualitatively, as the lifetime shortens at higher energies, the population proportion for higher $\varepsilon$ will increase resulting in a blue shift of the average energy content $k_B T_0$ is expected at sufficiently high incident light intensity. To elaborate, in contrast to the simplified three level system that Liu *et al*[12] has proposed, here a more generalized and realistic model based on the continuously excited hot carriers will be derived, with the average energy now given by,

$$k_B T_0 = \frac{\int_0^{E_0} d\varepsilon \cdot \varepsilon r(\varepsilon) g(\varepsilon) \dfrac{exp\left(-\dfrac{\beta\zeta(\varepsilon+\delta)^2}{P}\right)}{1 - exp\left(-\dfrac{\beta\zeta(\varepsilon+\delta)^2}{P}\right)}}{\int_0^{E_0} d\varepsilon \cdot r(\varepsilon) g(\varepsilon) \dfrac{exp\left(-\dfrac{\beta\zeta(\varepsilon+\delta)^2}{P}\right)}{1 - exp\left(-\dfrac{\beta\zeta(\varepsilon+\delta)^2}{P}\right)}} \qquad (S.11)$$

where $g(\varepsilon)$ is the energy level degeneracy at $\varepsilon$ representing the density of states, which is commonly used in the calculation for carrier energy via Fermi integral in solid state physics, $\delta$ is the damping energy in order to avoid the singularity and is typically around $250 meV$, $E_0$ is the cutoff energy for the generated hot carriers and is equal to the incident photon energy ($\sim 1.58 eV$ in our case)[13]. Detailed modeling of optically driven hot carrier temperature $T_0$ requires the inclusion of metal's band structure, while the focus here is on the average hot carrier distribution energy and its power dependence. Hence, for general arguments, $r(\varepsilon)$ and $g(\varepsilon)$ are approximated to be energy independent uniform value $r_0$ and $g_0$ over the energy range from $\varepsilon = 0$ to $E_0$, but it can be easily generalized to a more complicated case. This assumption gives

$$k_B T_0 = \frac{\dfrac{P}{2\beta\zeta}\left[ln\left(1 - exp\left(\dfrac{\beta\zeta E_0^2}{P}\right)\right) - ln\left(1 - exp\left(\dfrac{\beta\zeta \delta^2}{P}\right)\right)\right] - \dfrac{1}{2}(E_0^2 - \delta^2)}{\int_0^{E_0} d\varepsilon \cdot \dfrac{exp\left(-\dfrac{\beta\zeta(\varepsilon+\delta)^2}{P}\right)}{1 - exp\left(-\dfrac{\beta\zeta(\varepsilon+\delta)^2}{P}\right)}} \qquad (S.12)$$

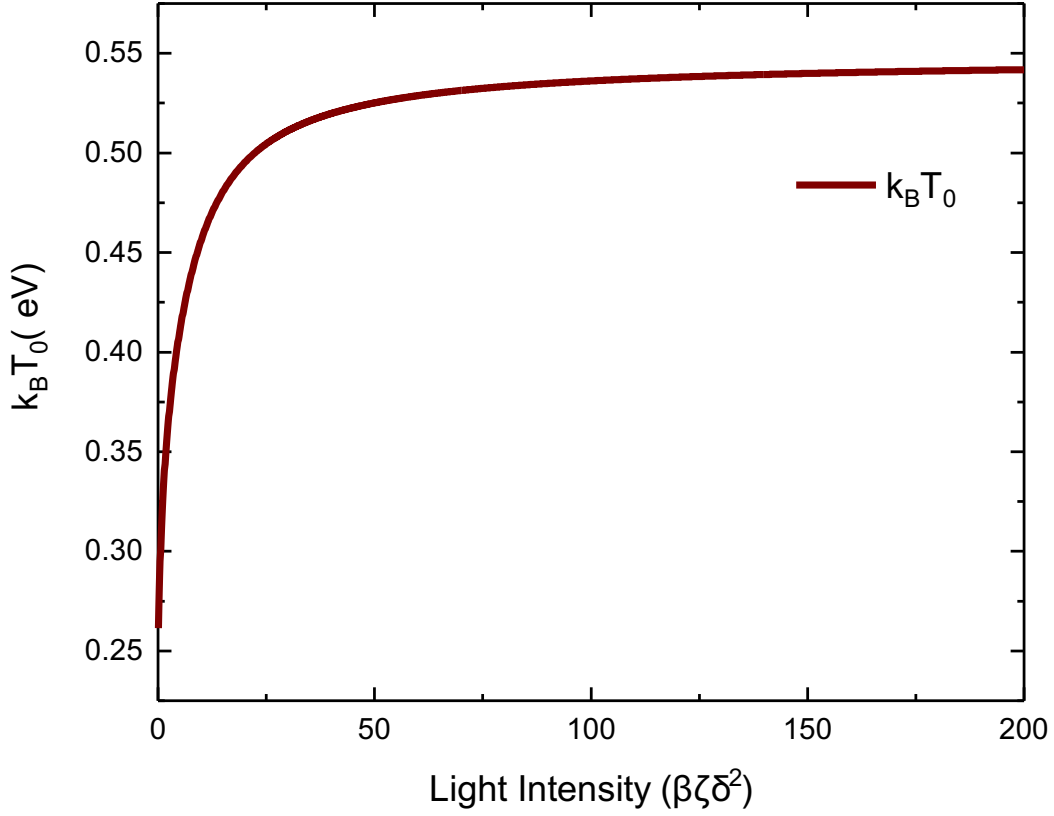

**Figure S6.** Numerically calculated optically driven hot carrier temperature. Calculated $k_B T_0$ varies as a function of the parameterized incident light intensity (in the units of $\beta\zeta\delta^2$). The energy blue shifted and saturates around $0.54 eV$.

The indefinite integral in the denominator is not analytical, and needs to be calculated numerically. Figure S6 depicts the numerically calculated $k_B T_0$ as a function of the parameterized incident light intensity (in the units of $\beta\zeta\delta^2$), showing qualitatively similar blue shift and saturation behavior with Liu *et al*[12]. The optically driven hot carrier temperature $T_0$ raises rapidly and saturate to around $0.54 eV$ at sufficiently large light intensity.

For a 100nm wide, 600nm long nanowire investigated here, optically excited hot carrier distribution will mainly be dominated by a strong peak at $\varepsilon = 0.2 eV$ above the Fermi energy[14]. Inspired by this work and for the sake of simplicity, here we consider a two-level model in which level $|0\rangle$ and $|1\rangle$ (with arbitrary degeneracies) represent a state essentially at the Fermi energy which will finally decay to become the normal carriers and a distinct hot electron continuum denoting the excited energy peak at $0.2\ eV$. Hence, from the aforementioned derivation, the hot carrier population at each energy level $p_0(P)$ and $p_1(P)$ are given by

$$p_0(P) = \frac{exp\left(-\frac{\beta\zeta\delta^2}{P}\right)}{1 - exp\left(-\frac{\beta\zeta\delta^2}{P}\right)}, \quad p_1(P) = \frac{exp\left(-\frac{\beta\zeta(\delta + E_1)^2}{P}\right)}{1 - exp\left(-\frac{\beta\zeta(\delta + E_1)^2}{P}\right)} \quad (S.13)$$

where $E_1 = 0.2\ eV$ is the energy peak. We can then obtain the steady-state time average of the energy content

$$\bar{E}_{hc}(P) = \frac{p_0 E_0 + p_1 E_1}{p_0 + p_1} \quad (S.14)$$

from which the power dependent effective temperature of the hot carriers can then be obtained via $k_B T_0 = \bar{E}_{hc}(P)$. In performing our calculation (Fig. 4A and 4B), we used the following parameters: the photon energy is 1.58eV (785 nm), the absorption cross-section $\sigma$ of the nanostructure is taken to be equal to the surface area of the nanowire, which is $0.06\ \mu m^2$, the hot carrier relaxation time scale $\tau_1$ at $E_1$ is taken as 200fs which is typical for the carrier relaxation due to electron-electron interaction in gold nanostructures[12,15].

In the presence of simultaneous optical and electrical excitation, the Boltzmann-like form of the emission suggests that it still makes sense to define an effective temperature for a steady-state hot carrier tail. This can be estimated as follows[10]. Let us assume that the optical pumping generates an effective temperature, $T_{PL}$ (as determined above) which is *independent* of the contribution of the electron viscosity. The optical pumping is then a source of power in the junction, $I_{PL} = \gamma T_{PL}^2$, where $\gamma$ is some heat coefficient that depends on the junction geometry. The contribution from the electron viscosity is instead $I_{EL} = \gamma T_{EL}^2$, where $T_{EL}$ was estimated according to Eqs. (S.7) and (S.8), and we have assumed the same heat coefficient $\gamma$ because the geometric properties of the junction are the same for all heat sources. The total power carried away from the junction is then $I_{EPL} \equiv \gamma T_{EPL}^2 = \gamma T_{EL}^2 + \gamma T_{PL}^2$. From this relation we can then estimate the effective temperature $T_{EPL}$ under the combined external stimuli as,

$$T_{EPL} = \sqrt{T_{EL}^2 + T_{PL}^2} \tag{S.15}$$

where the $T_{EL}$ is the effective temperature under pure electrical stimulation, $T_{PL}$ is the effective temperature under optical pumping only. We stress that this equation holds only when the energy pumped by the laser creates an incoherent temperature contribution $T_{PL}$ which is independent of the effective temperature due to the current-carrying electrons scattering on plasmons.

Combining Eq. (S.3) and Eq. (S.15) together and with the help of the Heaviside step function $\theta(x)$, the generalized formula of EPL/EL/PL can then be constructed

$$U^{EPL/EL/PL} \propto [\theta(-I) + \theta(I)I^\alpha] \cdot [\theta(-P) + \theta(P)P^\beta]\rho(\omega)\hbar\omega \cdot exp\left(-\frac{\hbar\omega}{k_B T_{eff}^{EPL/EL/PL}}\right) \tag{S.16}$$

where $T_{eff}$ here denotes $T_{eff}^{EPL}$, $T_{eff}^{EL}$ and $T_{eff}^{PL}$ in each case, which are Eq. (S.7), Eq. (S.14) and Eq. (S.15) respectively.

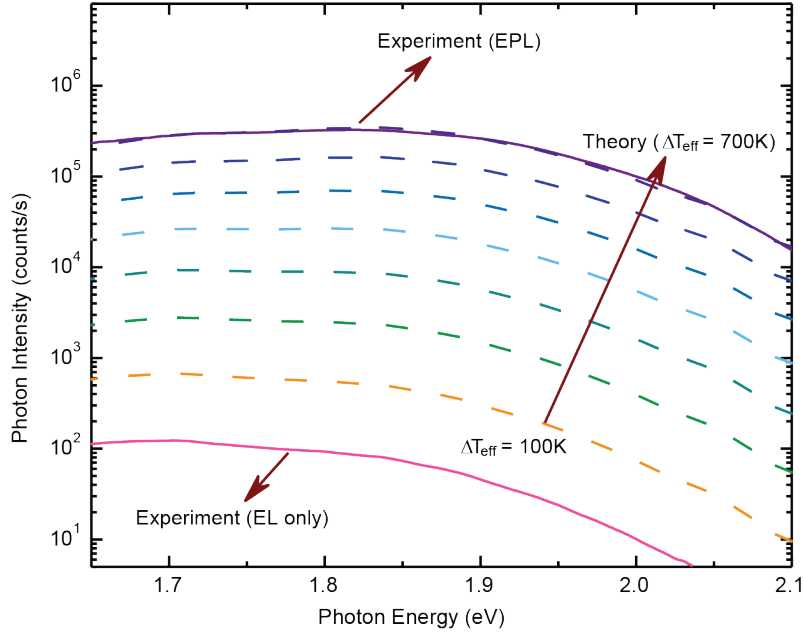

**Figure S7.** Numerically calculated EPL spectrum using the theoretical model plotted on logarithmic scale. Every dotted line represents an effective temperature rise of 100K, with the top dotted line excellently matching the EPL curve at $\Delta T_{eff} = 700K$. The normalized curves in this plot are shown in Fig. 4C to demonstrate more clearly the blue shift behavior when increasing the effective temperature.

Once the prefactor in Eq. (S. 16), $\rho(\omega)$, and the effective temperature dependence under different external stimuli have been extracted from the experiment, this equation can be used to reproduce the EPL spectrum based on the measured EL spectrum. To elaborate, simply by inserting the effective temperature under respective external stimuli and combining other terms in the expression, one can calculate the spectrum expected under different effective temperatures. When the effective temperature rises, a blue shift in the spectrum will occur due to the energy dependent Boltzmann factor (Fig. 4C), though this is less obvious when plotted in logarithmic scale (Fig. S7). In addition, the amplitude of the spectrum also grows since the temperature is in an exponential expression, which can be clearly seen in Fig. S8 where each dashed line corresponds to an effective temperature rise of an additional 100 K. When the inserted effective temperature is equal to $T_{EPL}$, a near-perfect match between the calculated the spectrum and the measured EPL can then be achieved, further validating our theoretical model.

Furthermore, the generalized formula for EPL/EL/PL we derived enables the numerical estimation of the enhancement ratio via Eq. (S15) and Eq. (S16). Here by calculating the enhancement ratio using the definition in Eq. (1) we are hoping to show the cooperative interaction between electrical and optical excitation, where the optically induced effective temperature $T_{PL}$ provides a nonlinear increase in the exponential factor, resulting in the giant enhancement. Hence the synergistic enhancement will be governed by the relative value of $T_{EL}$ and $T_{PL}$. Intuitively, in the high bias limit, optically induced effective temperature increase $(T_{EPL} - T_{EL})$ is small compared to the large $T_{EL}$ value, leading to a decreasing behavior for the enhancement ratio as a function of bias. In this regime the enhancement ratio is mainly dominated by the electroluminescence ( $U^{EPL}/[U^{PL} + U^{EL}] \sim U^{EPL}/U^{EL}$ ). By contrast, at the low bias limit, where $(T_{EPL} - T_{EL}) \sim T_{PL}$, photoluminescence dominates the enhancement ratio ( $U^{EPL}/[U^{PL} + U^{EL}] \sim U^{EPL}/U^{PL}$ ), which will increase when cranking up the bias since $U^{EPL}$ grows rapidly compared to the slow varying sum of $U^{PL} + U^{EL}$. Hence, there must be an optimum bias with the largest enhancement ratio. This bias value can be physically interpreted as the energy scale of the electrically driven hot carries being comparable to the optically excited ones. Under this optimum bias, further increasing the laser power will result in the enhancement ratio rapidly increasing and then asymptotically approaching a fixed value given that both $U^{EPL}$ and $U^{PL}$ are slightly superlinear in optical power (Eq. (S.16)). Hence, the best condition reaches under this device dependent optimum bias and a moderate optical power where the enhancement ratio are close to the fixed value without damaging the junction.